\documentclass[twocolumn]{aastex62}

\graphicspath{{./}{figures/}}

\received{}
\revised{}
\accepted{}

\shorttitle{ASAGAO: NEAR-IR DARK FAINT ALMA SOURCES}
\shortauthors{Yamaguchi et al.}

\begin{document}

\title{ALMA TWENTY-SIX ARCMIN$^2$ SURVEY OF GOODS-S AT ONE-MILLIMETER (ASAGAO): NEAR-INFRARED-DARK FAINT ALMA SOURCES}

\correspondingauthor{Yuki Yamaguchi}
\email{yyamaguchi@ioa.s.u-tokyo.ac.jp}

\author[0000-0002-0786-7307]{Yuki Yamaguchi}
\affil{Institute of Astronomy, Graduate School of Science, The University of Tokyo, 2-21-1 Osawa, Mitaka, Tokyo 181-0015, Japan}

\author{Kotaro Kohno}
\affiliation{Institute of Astronomy, Graduate School of Science, The University of Tokyo, 2-21-1 Osawa, Mitaka, Tokyo 181-0015, Japan}
\affiliation{Research Center for the Early Universe, Graduate School of Science, The University of Tokyo, 7-3-1, Hongo, Bunkyo, Tokyo 113-0033, Japan}

\author{Bunyo Hatsukade}
\affiliation{Institute of Astronomy, Graduate School of Science, The University of Tokyo, 2-21-1 Osawa, Mitaka, Tokyo 181-0015, Japan}

\author{Tao Wang}
\affiliation{Institute of Astronomy, Graduate School of Science, The University of Tokyo, 2-21-1 Osawa, Mitaka, Tokyo 181-0015, Japan}
\affiliation{National Astronomical Observatory of Japan, 2-21-1 Osawa, Mitaka, Tokyo 181-8588, Japan}

\author{Yuki Yoshimura}
\affiliation{Institute of Astronomy, Graduate School of Science, The University of Tokyo, 2-21-1 Osawa, Mitaka, Tokyo 181-0015, Japan}

\author{Yiping Ao}
\affiliation{Purple Mountain Observatory and Key Laboratory for Radio Astronomy, Chinese Academy of Sciences, 8 Yuanhua Road, Nanjing 210034, China}

\author{Karina I.~ Caputi}
\affiliation{Kapteyn Astronomical Institute, University of Groningen, P.O. Box 800, 9700AV Groningen, The Netherlands}

\author{James S.~Dunlop}
\affiliation{Institute for Astronomy, University of Edinburgh, Royal Observatory, Edinburgh EH9 3HJ, UK}

\author{Eiichi Egami}
\affiliation{Steward Observatory, University of Arizona, 933 North Cherry Avenue, Tucson, AZ 85721, USA}

\author{Daniel Espada}
\affiliation{National Astronomical Observatory of Japan, 2-21-1 Osawa, Mitaka, Tokyo 181-8588, Japan}
\affiliation{Department of Astronomical Science, SOKENDAI (The Graduate University of Advanced Studies), 2-21-1 Osawa, Mitaka, Tokyo 181-8588, Japan}

\author{Seiji Fujimoto}
\affiliation{Institute for Cosmic Ray Research, The University of Tokyo, Kashiwa, Chiba 277-8582, Japan}

\author{Natsuki H.~Hayatsu}
\affiliation{Department of Physics, Graduate School of Science, The University of Tokyo, 7-3-1 Hongo, Bunkyo, Tokyo 113-0033, Japan}
\affiliation{European Southern Observatory, Karl-Schwarzschild-Str.~2, D-85748 Garching, Germany}

\author{Rob J.~Ivison}
\affiliation{Institute for Astronomy, University of Edinburgh, Royal Observatory, Edinburgh EH9 3HJ, UK}
\affiliation{European Southern Observatory, Karl-Schwarzschild-Str.~2, D-85748 Garching, Germany}

\author{Tadayuki Kodama}
\affiliation{Astronomical Institute, Tohoku University, 6-3 Aramaki, Aoba, Sendai, Miyagi 980-8578, Japan}

\author{Haruka Kusakabe}
\affiliation{Department of Astronomy, Graduate School of Science, The University of Tokyo, 7-3-1 Hongo, Bunkyo, Tokyo 113-0033, Japan}

\author{Tohru Nagao}
\affiliation{Research Center for Space and Cosmic Evolution, Ehime University, 2-5 Bunkyo-cho, Matsuyama, Ehime 790-8577, Japan}

\author{Masami Ouchi}
\affiliation{Institute for Cosmic Ray Research, The University of Tokyo, Kashiwa, Chiba 277-8582, Japan}
\affiliation{Kavli Institute for the Physics and Mathematics of the Universe (Kavli IPMU), WPI, The University of Tokyo, Kashiwa, Chiba 277-8583, Japan}

\author{Wiphu Rujopakarn}
\affiliation{Kavli Institute for the Physics and Mathematics of the Universe (Kavli IPMU), WPI, The University of Tokyo, Kashiwa, Chiba 277-8583, Japan}
\affiliation{Department of Physics, Faculty of Science, Chulalongkorn University, 254 Phayathai Road, Pathumwan, Bangkok 10330, Thailand}
\affiliation{ National Astronomical Research Institute of Thailand (Public Organization), Don Kaeo, Mae Rim, Chiang Mai 50180, Thailand}

\author{Ken-ichi Tadaki}
\affiliation{National Astronomical Observatory of Japan, 2-21-1 Osawa, Mitaka, Tokyo 181-8588, Japan}

\author{Yoichi Tamura}
\affiliation{Division of Particle and Astrophysical Science, Nagoya University, Furocho, Chikusa, Nagoya 464-8602, Japan}

\author{Yoshihiro Ueda}
\affiliation{Department of Astronomy, Kyoto University, Kyoto 606-8502, Japan}

\author{Hideki Umehata}
\affiliation{RIKEN Cluster for Pioneering Research, 2-1 Hirosawa, Wako-shi, Saitama 351-0198, Japan}
\affiliation{Institute of Astronomy, Graduate School of Science, The University of Tokyo, 2-21-1 Osawa, Mitaka, Tokyo 181-0015, Japan}

\author{Wei-Hao Wang}
\affiliation{Academia Sinica Institute of Astronomy and Astrophysics (ASIAA), No.~1, Sec.~4, Roosevelt Rd., Taipei 10617, Taiwan}
\affiliation{Canada-France-Hawaii Telescope (CFHT), 65-1238 Mamalahoa Hwy., Kamuela, HI 96743, USA}

\author{Min S.~Yun}
\affiliation{Department of Astronomy, University of Massachusetts, Amherst, MA 01003, USA}



\begin{abstract}
We report detections of two 1.2 mm continuum sources ($S_\mathrm{1.2mm}$ $\sim$ 0.6 mJy) without any counterparts in the deep $H$- and/or $K$-band image (i.e., $K$-band magnitude $\gtrsim$ 26 mag). These {near-infrared-dark faint millimeter sources} are uncovered by ASAGAO, a deep and wide-field ($\simeq$ 26 arcmin$^2$) Atacama Large Millimeter/submillimeter Array (ALMA) 1.2 mm survey. One has a red IRAC (3.6 and 4.5 $\mu$m) counterpart, and the other has been independently detected at 850 and 870 $\mu$m using SCUBA2 and ALMA Band 7, respectively. Their optical to radio spectral energy distributions indicate that they can lie at $z\gtrsim$ 3--5 and can be in the early phase of massive galaxy formation. Their contribution to the cosmic star formation rate density is estimated to be $\sim$ 1 $\times$ 10$^{-3}$ $M_\odot$ yr$^{-1}$ Mpc$^{-3}$ if they lie somewhere in the redshift range of $z\sim$ 3--5. This value can be consistent with, or greater than that of bright submillimeter galaxies ($S_\mathrm{870\mu m}>$ 4.2 mJy) at $z\sim$ 3--5. We also uncover 3 more candidate near-infrared-dark faint ALMA sources without any counterparts ($S_\mathrm{1.2mm}$ $\sim$ 0.45--0.86 mJy). These results show that an unbiased ALMA survey can reveal the dust-obscured star formation activities, which were missed in previous deep optical/near-infrared surveys.
\end{abstract}



\section{Introduction} \label{sec:intro}

The advent of the Atacama Large Millimeter/sub-millimeter Array (ALMA), which offers high sensitivity and angular resolution capabilities, has enabled us to uncover faint (sub-)millimeter populations (observed flux densities, $S_\mathrm{obs}$ $\simeq$ 0.1--1 mJy, corresponding to infrared luminosity of $L_\mathrm{IR}\lesssim$ 10$^{12}$ $L_\odot$\footnote{Rest-frame 8--1000 $\mu$m.}). Recently, several blind surveys using ALMA have been performed in the SXDF \citep[e.g.,][]{tadaki2015, kohno2016, hatsukade2016, yamaguchi2016, wangwh2016} and the GOODS-S field \citep[Yamaguchi et al.~submitted to ApJ]{aravena2016, dunlop2017, ueda2018, fujimoto2018, franco2018, hatsukade2018} {to detect and characterize the faint (sub-)millimeter galaxies (hereafter, faint SMGs). These studies suggest that they are primarily ``typical'' or``the main-sequence" star-forming galaxies at $z$ = 1--4 \citep[e.g.,][Yamaguchi et al.~submitted to ApJ]{dacunha2015, aravena2016, yamaguchi2016, dunlop2017}, based on the cross-matching of the ALMA-selected sources and optical to near-infrared-selected sources with reliable photometric redshifts and stellar mass estimates. }

{Here, we focus on the ALMA-selected galaxies which are not well characterized by such a cross-matching technique, i.e., faint SMGs without significant counterpart seen in the optical and near-infrared (near-IR) wavelengths. The existence of optical/near-IR-dark SMGs have already reported by using pre-ALMA interferometers \citep[e.g.,][]{yun2008, wang2009, tamura2010}.} In the ALMA era, {\citet{simpson2014} found that a significant fraction (17 out of 96) of the bright ALMA sources in ECDF-S, which are originally selected by the LABOCA on APEX survey at 870 $\mu$m, are too faint in the optical/near-IR bands to obtain reliable constraints on their photometric redshift, arguing that such ``near-IR-dark'' SMGs tend to lie at higher redshift than the typical SMGs based on the \textit{Herschel} stacking. Similarly, ALMA follow-up observations of SCUBA2-selected SMGs in UDS revealed that 4 bright ALMA sources out of 23 does not have significant near-IR counterparts \citep{simpson2015}. And in fact, such trend extends to the faint SMGs purely selected by ALMA. For instance, }\cite{fujimoto2016} suggest that $\simeq$ 40\% of faint ALMA sources {($S_\mathrm{1.2mm}$ = 0.02--1 mJy) uncovered in the ALMA archival images of various fields (the total coverage is $\sim$9 arcmin$^2$)} have no counterparts at optical/near-IR wavelengths {(the 5$\sigma$ limiting magnitude of $\sim$ 27--28 mag at optical wavelengths and $\sim$ 25--26 mag at near-IR wavelengths)}. \cite{yamaguchi2016} find that one out of five ALMA sources {in the 2 arcmin$^2$ survey of SXDF ($S_\mathrm{1.1mm}$ = 0.54--2.0 mJy)} are faint at $H$-band ($\simeq$ 25.3 mag) and not detected at wavelengths shorter than $\sim$ 1.3 $\mu$m. {All these studies strongly motivate us to conduct a systematic search for near-IR-dark faint SMGs in the fields where the deepest near-IR images to date are available.}

In this paper, we report detections of near-IR-dark, faint ALMA sources {($S_\mathrm{1.2mm}$ = 0.45--0.86 mJy), which do not have} any significant counterparts {in the ultra-deep $H$- and/or $K$-band images}, based on the ALMA twenty Six Arcmin$^2$ survey of GOODS-S At One-millimeter (ASAGAO; Project ID: 2015.1.00098.S, PI: K.~Kohno). This paper is structured as follows. Section \ref{sec:observation} presents ALMA observations and ALMA source identification. In section \ref{sec:K-drop}, we describe the properties of the near-IR-dark faint ALMA sources detected by ASAGAO. {Then, we put constraints on their physical properties such as redshifts and stellar masses in Section \ref{sec:properties}.} Finally, we estimate their contribution to the cosmic star formation rate density (SFRD) in the high redshift universe (Section \ref{sec:sfrd}). Throughout this paper, we assume a $\Lambda$ cold dark matter cosmology with $\Omega_\mathrm{M}$ = 0.3, $\Omega_\mathrm{\Lambda}$ = 0.7, and $H_0$ = 70 km s$^{-1}$ Mpc$^{-1}$. All magnitude are given according to the AB system. We adopt the Chabrier Initial Mass Function \citep[IMF;][]{chabrier2003} when necessary to compute the SFR in galaxies in this paper.

\section{ALMA source catalog and identifications of near-IR-dark faint ALMA candidates} \label{sec:observation}

\begin{deluxetable*}{ccccccccc}
\tabletypesize{\footnotesize}
\tablecaption{$K$-dropout ASAGAO sources \label{tab:Kdrop_source}}
\tablewidth{0pt}
\tablehead{
\colhead{ID} & \colhead{RA} & \colhead{Dec.} & \colhead{$S_\mathrm{ALMA}$} & \colhead{S/N$_\mathrm{peak}$} & \colhead{PB coverage\tablenotemark{a}} & \colhead{$z_\mathrm{3.6\mu m/1.2mm}$} &  \colhead{$z_\mathrm{5.0cm/1.2mm}$} & \colhead{counterpart?} \\
\colhead{(ASAGAO)} & \colhead{(deg.)} & \colhead{(deg.)} & \colhead{(mJy)}  & \colhead{} & \colhead{} & \colhead{} & \colhead{} & \colhead{}
}
\decimalcolnumbers
\startdata
17 & 53.206042 & $-$27.819166 & 0.564 $\pm$ 0.090 & 6.078 & 0.403 & 3.93$^{+0.43}_{-0.30}$  & $>$4.14 & Y \\
20 & 53.120445 & $-$27.742093 & 0.614 $\pm$ 0.109 & 5.565 & 0.317 & $>$5.52 & $>$4.39 & Y \\
\hline
22 & 53.171662 & $-$27.817153 & 0.612 $\pm$ 0.101 & 5.446 & 0.483 & $>$5.41 & $>$4.26 & -- \\
24 & 53.183284 & $-$27.755207 & 0.446 $\pm$ 0.082 & 5.022 & 0.572 & $>$4.48 & $>$3.70 & -- \\
25 & 53.201002 & $-$27.789483 & 0.858 $\pm$ 0.223 & 5.020 & 0.438 & $>$5.92 & $>$4.93 & -- \\
\enddata
\tablenotetext{\textrm{a}}{{The primary beam (PB) coverage values in Table \ref{tab:Kdrop_source} look smaller than typical values ($>$0.5), but this does not mean that these are sources outside nominal FoVs; this is simply caused by the non-uniform PB coverage of the ASAGAO final map (Figure \ref{fig:ASAGAO_map}, right panel), which was produced by combining three different ALMA programs including the HUDF data \citep{dunlop2017}, and the GOODS-S ALMA data \citep{franco2018} as well as the ASAGAO data. If we exclude the HUDF data, which cause the non-uniformity (the orange dashed region in Figure \ref{fig:ASAGAO_map}), the PB coverage value of ID17, ID20, ID22, ID24, and ID25 is 0.62, 0.48, 0.79, 0.97, and 0.69, respectively.}}
\tablecomments{(1) ASAGAO ID. (2) Right ascension (J2000). (3) Declination (J2000). (4) Spatial integrated flux density (de-boosted). (5) Peak S/N. {(6) The PB coverage at the position of $K$-dropout ASAGAO sources (see Figure \ref{fig:ASAGAO_map}). (7) and (8) are the photometric redshifts estimated by the flux ratios between 3.6 $\mu$m and 1.2 mm and 5.0 cm and 1.2 mm, respectively. Here, we assume the average SED of ALESS sources with $A_\mathrm{V}>$ 3.0 (see Section \ref{sec:properties}). (9) Based on the cross-matching with catalogs by \citet{ashby2015} and \citet{cowie2018}; ``Y'' is assigned if $K$-dropout ASAGAO sources have counterparts at \textit{Spitzer}/IRAC, JCMT/SCUBA2, or ALMA Band 7. As discussed in Section \ref{sec:K-drop}, ID17 and ID20 are secure detections, and the rest of three are treated as rather tentative.}}
\end{deluxetable*}

{We examined 25 secure ALMA sources with signal-to-noise ratio (S/N) $>$ 5 in the 26 arcmin$^2$ map of the ASAGAO \citep{hatsukade2018} to search for near-IR dark faint ALMA sources. {Here we adopt peak S/N values, rather than the spatially integrated S/N values, to conduct source extraction.} The details of the ALMA observations and the source catalog creation are given in \citet{hatsukade2018}. Here we provide a brief overview.} The 26 arcmin$^2$ map of the ASAGAO field was obtained at 1.14 mm and 1.18 mm (two tunings) to cover a wider frequency range, whose central wavelength was 1.16 mm. {To obtain the best ALMA image of this field,} we also include ALMA archival data toward the same field (Project ID: 2015.1.00543.S, PI: D.~Elbaz and Project ID: 2012.1.00173.S, PI: J.~S.~Dunlop). {After adopting a 250-k$\lambda$ taper, which gives an optimal combination of the sensitivity and angular resolution,} the final map reached a typical rms noise of {30 $\mu$Jy beam$^{-1}$ at the central $\sim$ 4 arcmin$^2$ and $\sim$ 70 $\mu$Jy beam$^{-1}$ at the remaining area with the synthesized beam 0$^{\prime\prime}$.59 $\times$ 0$^{\prime\prime}$.53 (PA = $-$83$^\circ$). }

\begin{figure*}[ht!]
\gridline{\fig{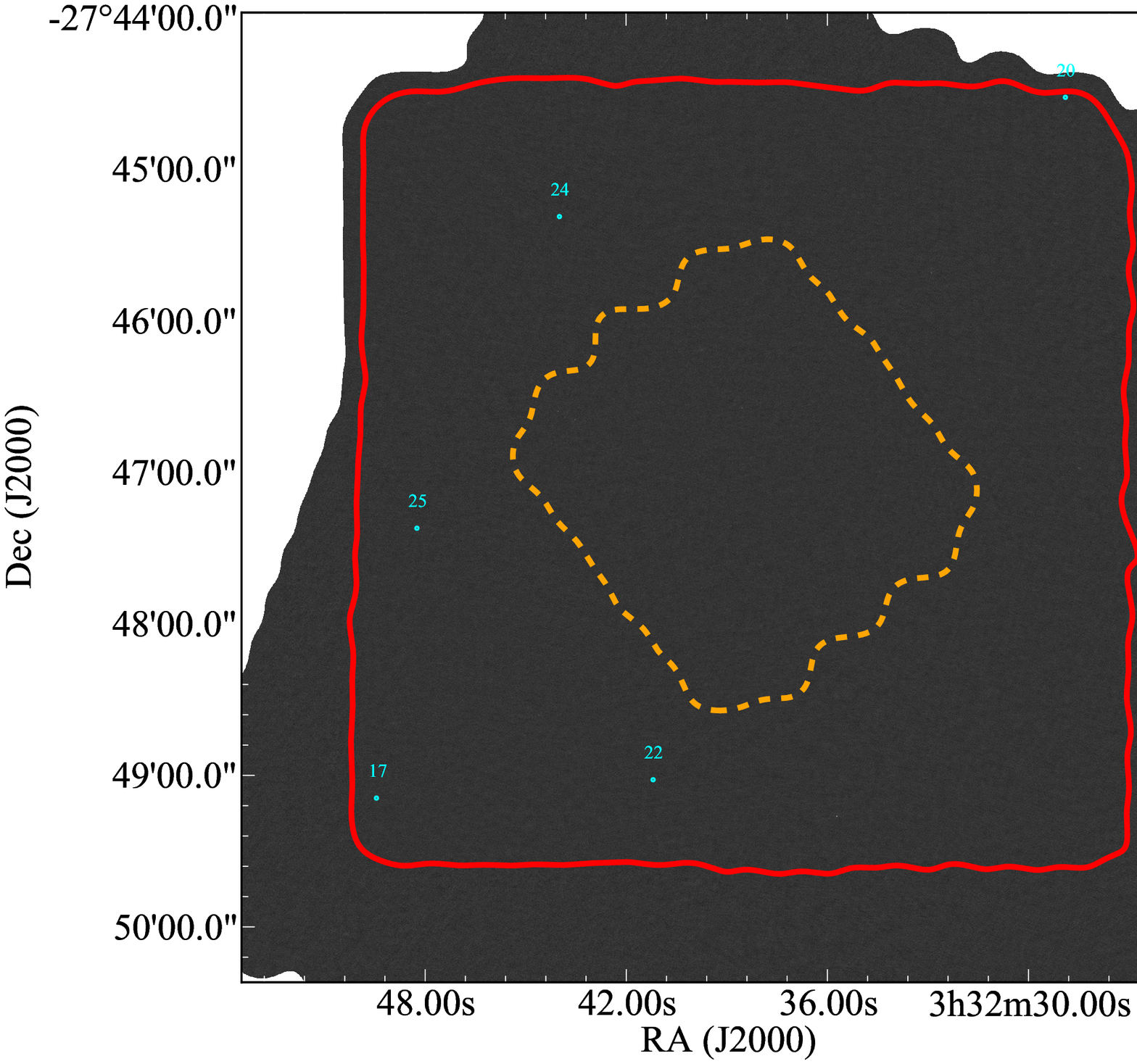}{0.455\textwidth}{}
\fig{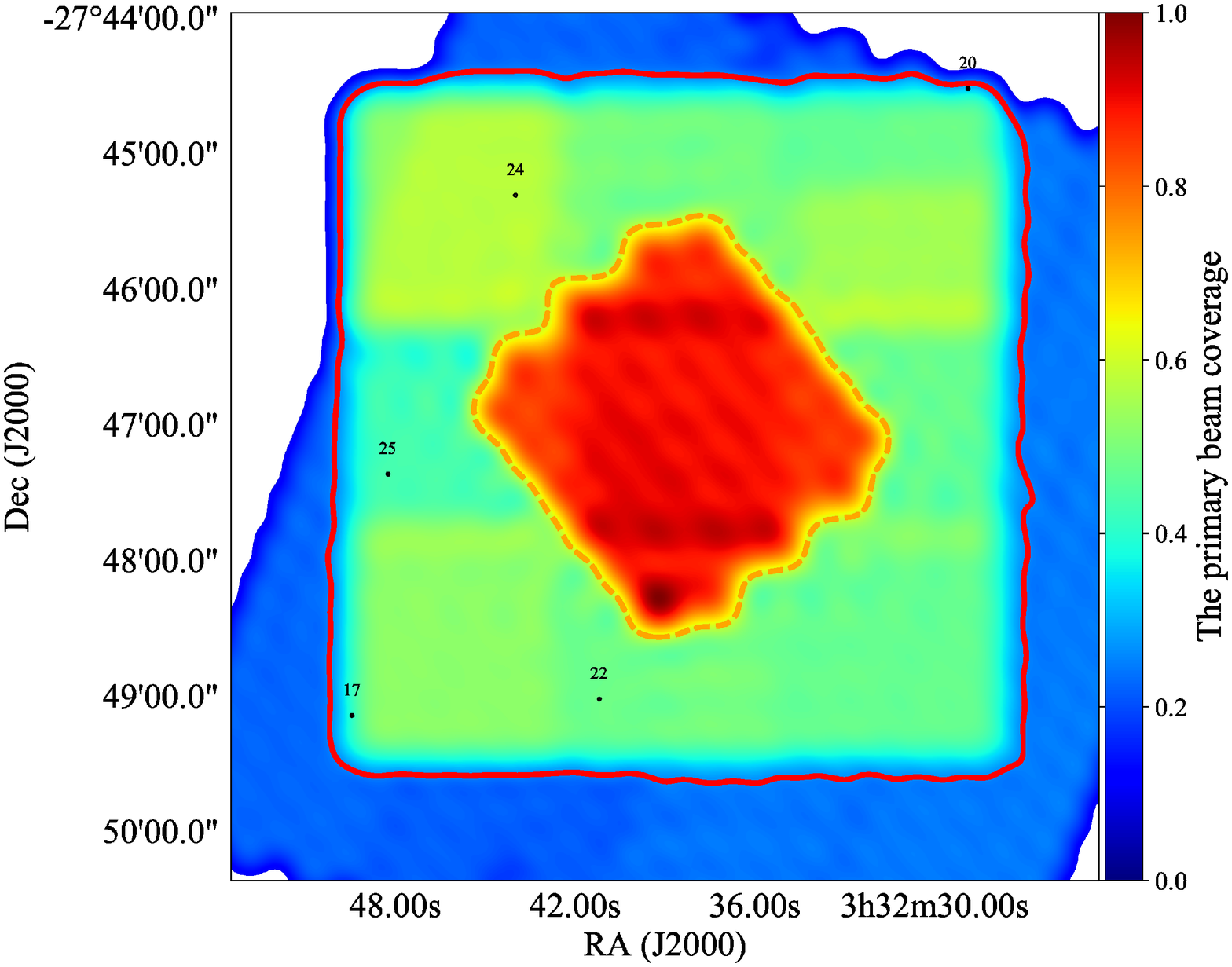}{0.50\textwidth}{}
}
\caption{{ASAGAO 1.2-mm continuum map of GOODS-S (the left panel) and the PB coverage map (the right panel). Here, we combined ASAGAO original data, HUDF data \citep{dunlop2017}, and a part of the GOODS-S ALMA data \citep{franco2018}. In this paper, we only consider the ASAGAO field indicated by the red solid line ($\sim$ 5$^{\prime}$ $\times$ 5$^{\prime}$). The orange dashed line indicates the area covered by \citet{dunlop2017}. The cyan and black circles in the right and left panel indicate 5 near-IR-dark ASAGAO candidates, respectively.}}
\label{fig:ASAGAO_map}
\end{figure*}

\begin{figure*}[ht!]
\epsscale{1.1}
\plotone{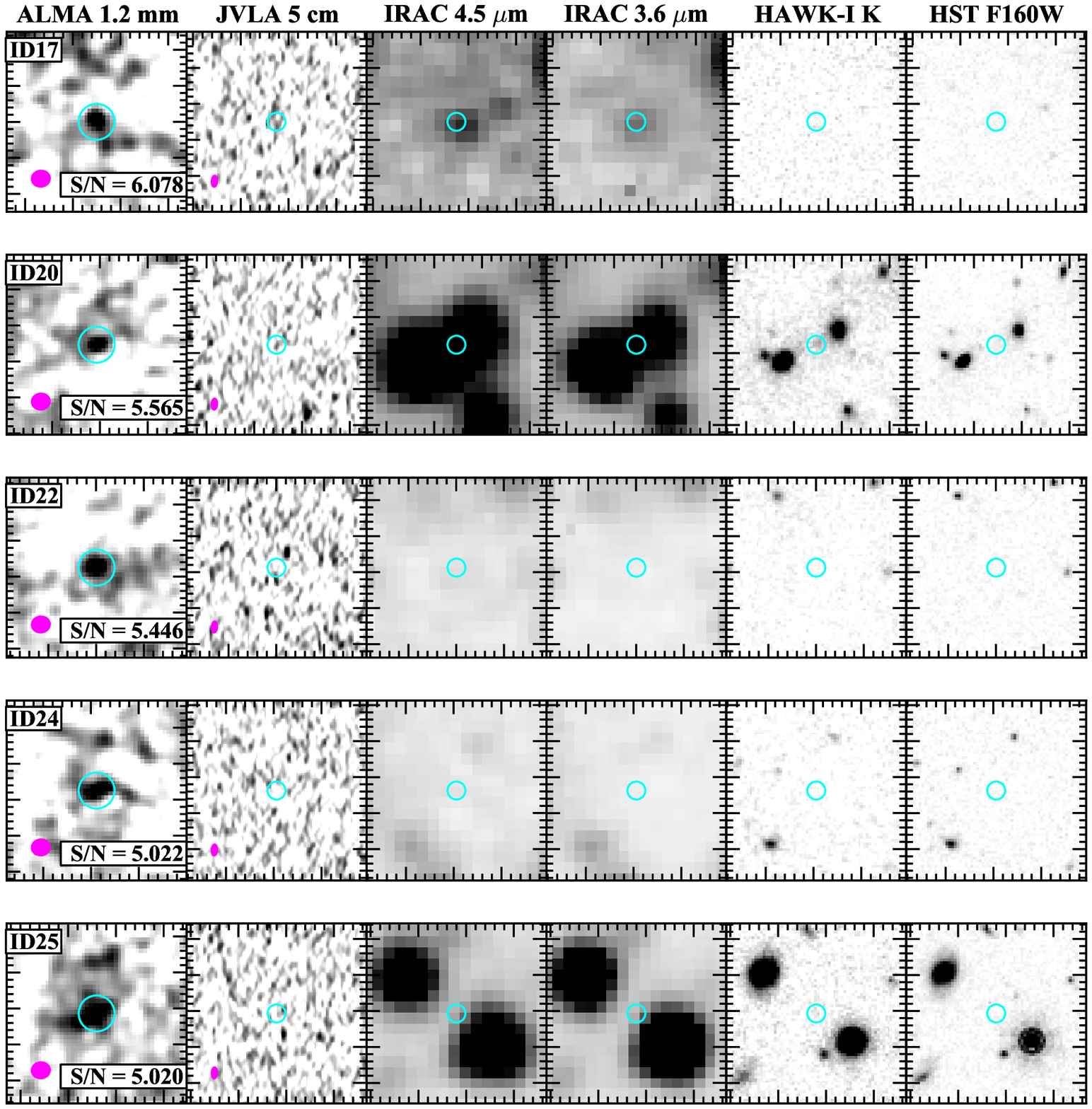}
\caption{Multi-wavelength images of ASAGAO {candidates} without $K$-band counterparts. From left to right: ALMA 1.2 mm (5$^{\prime\prime}$ $\times$ 5$^{\prime\prime}$), JVLA 5 cm, \textit{Spitzer} IRAC/4.5 $\mu$m, IRAC/3.6 $\mu$m, VLT HAWK-I/$K_s$, and HST WFC3/$F160W$ images (10$^{\prime\prime}$ $\times$ 10$^{\prime\prime}$). Cyan circles are 1$^{\prime\prime}$ apertures. The inserted S/N values are those of ALMA data. {The magenta symbol is the synthesized beam of ALMA and JVLA.}
\label{fig:stamp_no_counterpart}
}
\end{figure*}

Yamaguchi et al.~(submitted to ApJ) report that 20 of 25 sources candidates have been listed in $K$-band selected sources catalog by the \texttt{FourStar} galaxy evolution survey (ZFOURGE; \citealt{straatman2016}; the 5$\sigma$ limiting magnitude of $K_s$ = 26.0 mag at the 80\% completeness levels). {The ASAGAO sources are cross-matched against the ZFOURGE catalog, after correcting for a systematic offset with respect to the ALMA image ($-$0$^{\prime\prime}$.086 in right ascension and $+$0$^{\prime\prime}$.282 in declination), which is calibrated by the positions of stars in the Gaia Data Release 1 catalog \citet{gaia2016}. Here, we adopt the search radius $=$ 0$^{\prime\prime}$.5 for point-like sources, which is comparable with the synthesized beam of the final ALMA map. Considering the number of ZFOURGE sources within the ASAGAO field ($\sim$ 3,000), the likelihood of random coincidence is estimated to be 0.03 \citep[this likelihood is often called the $p$-value;][]{downes1986}. In the case that a counterpart is largely extended in the $K$-band image, we allow a larger positional offset, up to half-light radius of $K$-band emission.} However, we still have 5 {candidates} without ZFOURGE counterparts with S/N $>$ 5, which are undetected at $K$-band {(Figure \ref{fig:ASAGAO_map})}. We summarize the ASAGAO {candidates} without ZFOURGE counterparts  in Table \ref{tab:Kdrop_source}   and we show the multi-wavelength postage stamps of these 5 near-IR-dark faint ALMA {candidates} in Figure \ref{fig:stamp_no_counterpart}. 

We check the reliability of these near-IR-dark faint ALMA {candidates} using two independent methods. First, we apply the same source finding algorithm to the negative map in order to estimate the degree of contamination by spurious detections. The semi-analytical model by \citet{casey2018} suggests that the contamination rate is small in the range of S/N $>$ 5.0. There is no negative detections with S/N $>$ 5.2 to be compared with the 23 positive detections with S/N $>$ 5.2. In the 5.0 $<$ S/N $<$ 5.2 bin, we find one negative detections and two positive detections (i.e., ID24 and ID25; see Table \ref{tab:Kdrop_source}). Therefore, the negative fraction in the S/N bin is 0.5 (see also Figure 15 in \citealt{hatsukade2018}). Second, we split the ASAGAO visibilities into two polarization components (i.e., XX and YY polarization images) and create two XX and YY images, which are purely independent. With these two images, we find that all 5 {candidates} are detected with S/N $\sim$ 3--5 in both XX and YY. This is the behavior expected for $>5\sigma$ detections. {Based on these tests, we suggest that two highest S/N near-IR-dark ALMA sources seem to be secure, whereas remaining 3 sources may contain false detections. To test the reality of these sources further, we then consult with other deep images available in the next section.}

\section{$K$-dropout ASAGAO candidates} 
\label{sec:K-drop}
\begin{figure*}[ht!]
\plotone{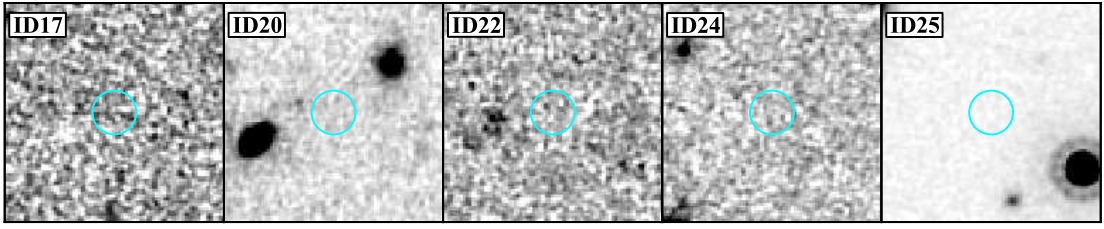}
\caption{{The stacked \textit{HST} images for K-dropout ASAGAO candidates (5$^{\prime\prime}$ $\times$ 5$^{\prime\prime}$). Cyan circles indicate ALMA positions (radius = 0$^{\prime\prime}$.5).}}
\label{fig:stacking}
\end{figure*}

In this section, we describe {ASAGAO candidates without ZFOURGE counterparts (hereafter, $K$-dropout ASAGAO candidates)} individually. {First, we perform a stacking analysis for each 5 $K$-dropout ASAGAO candidates using optical/near-IR images obtained by the 3D-\textit{HST} survey \citep{grogin2011, koekemoer2011,skelton2014}. This technique is often used to check reliability of extremely high redshift ($z\gtrsim$ 7) Lyman break galaxies \citep[e.g.,][]{bouwens2013}. We use the Advanced Camera for Survey \citep[ACS;][]{ford1998}/\textit{F435W}, \textit{F606W}, \textit{F775W}, \textit{F850LP}, \textit{F814W}, and the Wide Field Camera 3 \citep[WFC3;][]{kimble2008}/\textit{F125W}, \textit{F140W}, \textit{F160W}\footnote{These images are available at the 3D-\textit{HST} website; \url{https://3dhst.research.yale.edu/Data.php}}. In the stacking analysis, the Point Spread Functions (PSFs) of the \textit{HST} images are matched to the WFC3/\textit{F160W} image ($\simeq$ 0$^{\prime\prime}$.16). We show the results of the stacking in Figure \ref{fig:stacking}. We find no significant detections even in these ACS/WFC3 stacked images. Nevertheless, we find that two of the K-dropout ASAGAO candidates have independent detections in longer wavelengths as follows:}
\begin{itemize}
\item[--] {\textit{ID17}: This object is detected at 3.6 and 4.5 $\mu$m bands of \textit{Spitzer}/InfraRed Array Camera \citep[IRAC;][]{fazio2004} by the \textit{Spitzer}-Cosmic Assembly Deep Bear infrared Extragalactic Legacy Survey \citep[S-CANDELS; PI G.Fazio;][see Figure \ref{fig:stamp_no_counterpart}]{ashby2015}.  Its apparent magnitudes at 3.6 $\mu$m and 4.5 $\mu$m are 25.38 $\pm$ 0.30 and 25.00 $\pm$ 0.27 mag, respectively \citep{ashby2015}.}
\item[--] {\textit{ID20}: This object is detected at JCMT/SCUBA2 and ALMA Band 7 \citep[][]{cowie2018}. The observed flux density is 1.35 $\pm$ 0.24 mJy at 870 $\mu$m \citep[][]{cowie2018}. {This source is recognized as ID68 in \citet{cowie2018}}.}
\end{itemize}

{Considering the multi-wavelength information, two of the five $K$-dropout ASAGAO candidates with multi-wavelength counterparts (i.e., ID17 and ID20) must be real (secure detections), while we suggest that the rest of three candidates without multi-wavelength counterparts should remain ``candidates'', which shall be verified by further follow-up observations.}

\section{Physical properties} \label{sec:properties}

\begin{figure*}[ht!]
\plotone{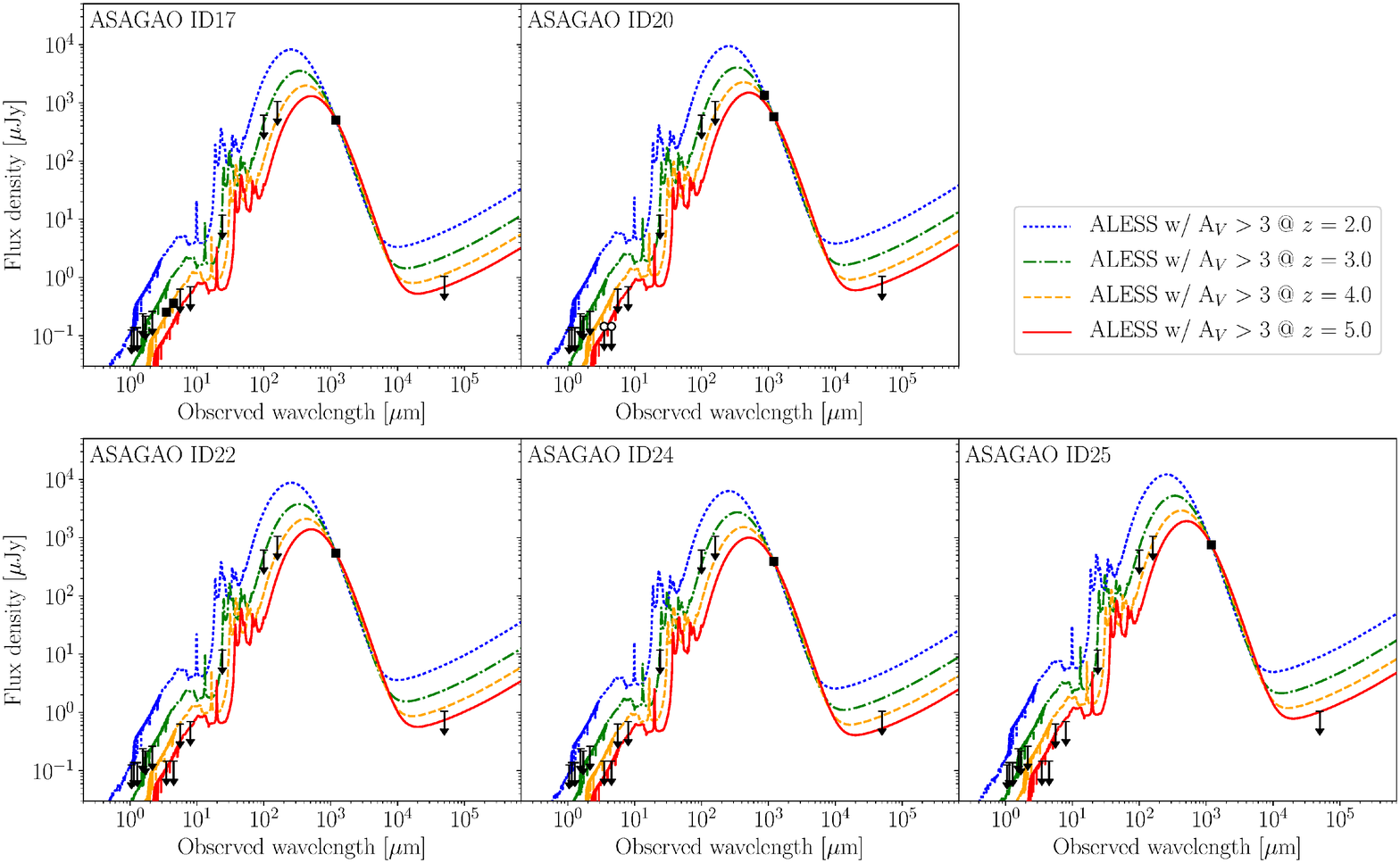}
\caption{Optical-to-radio SED of ASAGAO sources without $K$-band counterparts. Black arrow indicate 3$\sigma$ upper limits. From optical to far-IR upper limits except for IRAC 5.6 and 8.0 $\mu$m are listed in \citet{straatman2016}. The upper limits at IRAC 5.6 and 8.0 $\mu$m are presented in \citet{dickinson2003}. The radio image at 5 cm is obtained by JVLA (Rujopakarn et al.~in preparation). {For ID20, we also plot its ALMA Band 7 flux density. We have to note that we should not refer upper limits at \textit{Spitzer}/IRAC bands of ID20 (white open circles with upper limits) because of the contamination from nearby sources (Figure \ref{fig:stamp_no_counterpart}).} The blue dotted line, green dot-dashed line, orange dashed line, and red solid line indicate the average SED of ALESS sources with $A_\mathrm{V}$ $>$ 3.0 at $z$ = 2, 3, 4, and 5, respectively \citep{dacunha2015}. Note that these SEDs are scaled to their observed flux densities at ALMA wavelength.
\label{fig:SED_no_counterpart}
}
\end{figure*}
\begin{figure*}[ht!]
\epsscale{1.1}
\gridline{\fig{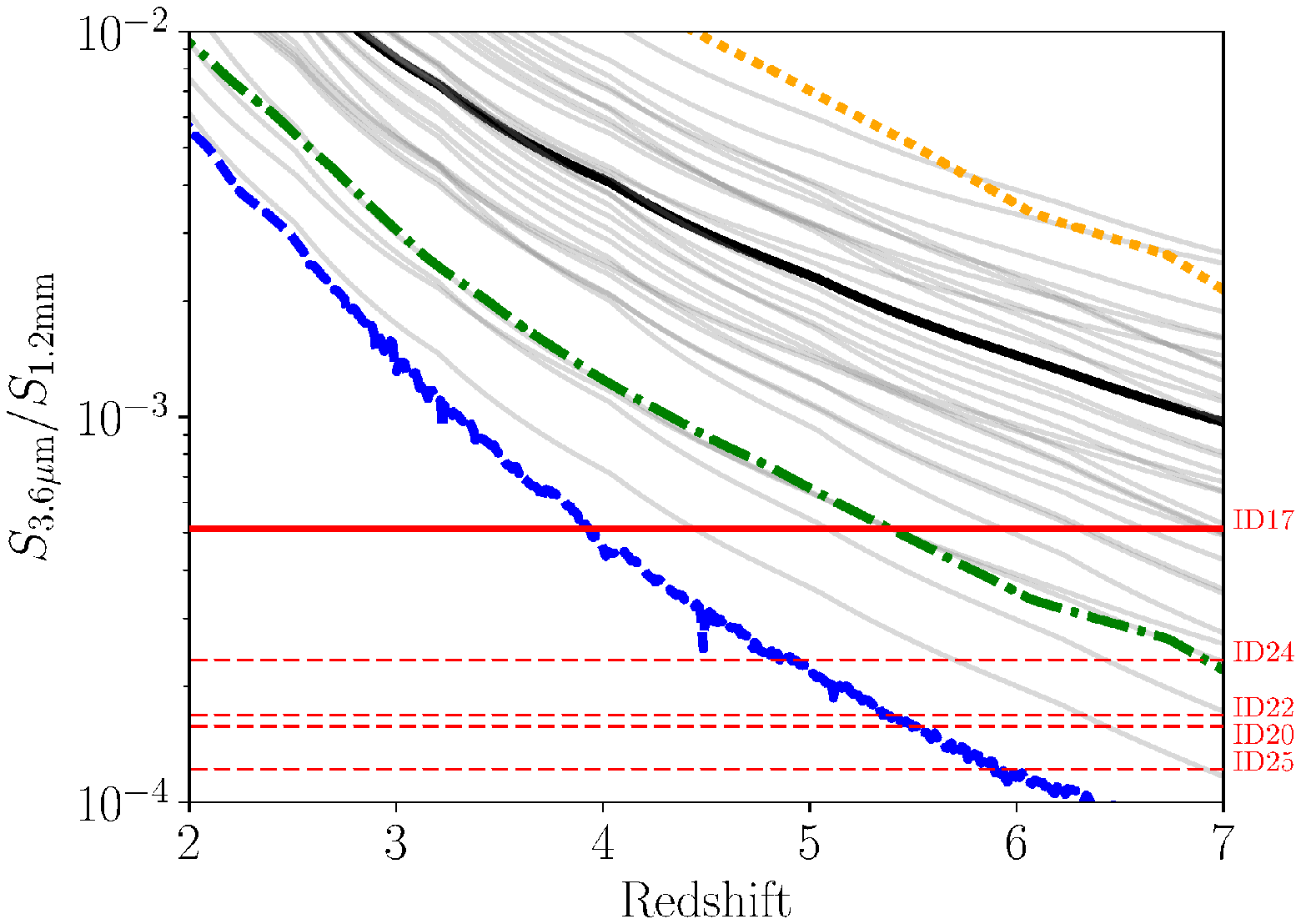}{0.50\textwidth}{}
          \fig{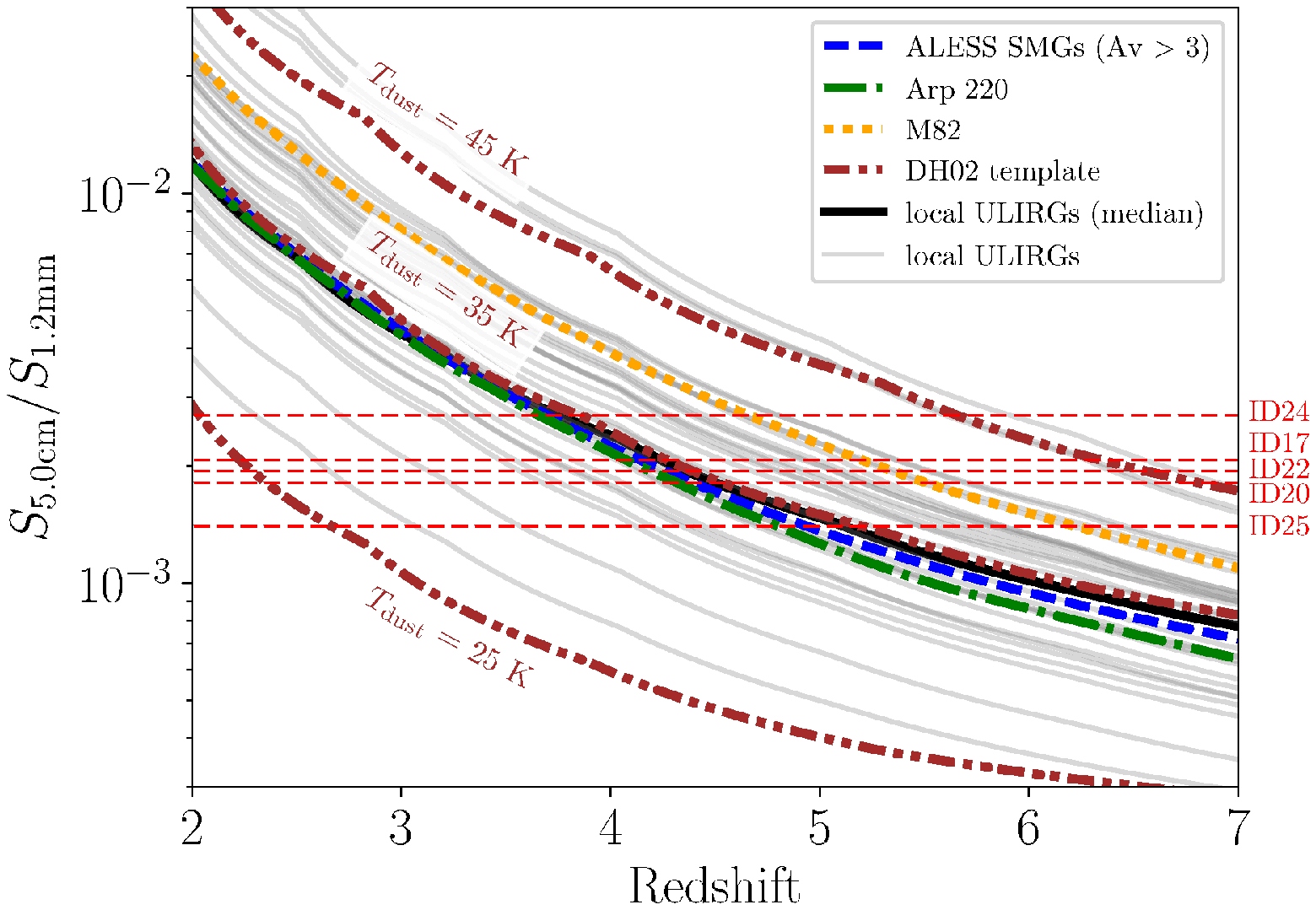}{0.50\textwidth}{}
}
\caption{{{Redshift dependence of $S_\mathrm{3.6\mu m}/S_\mathrm{1.2mm}$ (the left panel)} and $S_\mathrm{5cm}/S_\mathrm{1.2mm}$ (the right panel) flux ratios. The blue dashed line, green dot-dashed line, and orange dotted line indicates the average SED of ALESS sources with $A_\mathrm{V}>$ 3.0 \citep{dacunha2015}, Arp 220, and M82 \citep{silva1998}, respectively. Gray solid lines are local ULIRGs compiled by \citet{vega2008} and the black solid line is the median of these ULIRGs. Brown chain double dashed line in the right panel is SED templates of \citet{dale2002} with dust temperature, $T_\mathrm{dust}$ = 25, 35, and 45 K. {The horizontal red solid line in the left panel is the $S_\mathrm{3.6\mu m}/S_\mathrm{1.2mm}$ value of ID17.} Horizontal red lines in the left and right panels are the upper limit of $S_\mathrm{3.6\mu m}/S_\mathrm{1.2mm}$ (the left panel) and  $S_\mathrm{5cm}/S_\mathrm{1.2mm}$ (the right panel) flux ratios of ASAGAO $K$-dropout sources.}
\label{fig:radio-mm_redshift}
}
\end{figure*}

These extremely red colors can be reproduced by the high-redshift sources or highly-reddened low-redshift sources {\citep[e.g.,][]{caputi2012}}. We plot optical to radio Spectral Energy Distributions (SEDs) of these $K$-dropout ASAGAO {sources (including 3 candidates)} in Figure \ref{fig:SED_no_counterpart}. As a comparison, we also show the average SED of ALESS\footnote{The ALMA follow-up observation of the LABOCA Extended Chandra Deep Field South Survey \citep[e.g.,][]{hodge2013, dacunha2015, swinbank2014}} sources with visual extinction ($A_\mathrm{V}$) $>$ 3.0 (the reddest case; hereafter we call this SED as the average SED of ALESS SMGs) obtained by \citet[][]{dacunha2015}. As shown in Figure \ref{fig:SED_no_counterpart}, all sources can lie at $z\gtrsim$ 3--5, even though we assume the highly-reddened SED. {The relation between the flux ratio between 3.6 $\mu$m and 1.2 mm ($S_\mathrm{3.6\mu m}/S_\mathrm{1.2 mm}$; the left panel of Figure \ref{fig:radio-mm_redshift}) also prefer high redshift cases. For ID17 which is detected at 3.6 $\mu$m, the ratio indicates that it can lie at $z$ = 3.93$^{+0.43}_{-0.30}$, when we assume the average SED of ALESS sources. The redshift error is attributed to the error in the $S_\mathrm{3.6\mu m}/S_\mathrm{1.2 mm}$ ratio. On the other hand, as shown in the left panel of Figure \ref{fig:radio-mm_redshift}, variation between SEDs are quite large and some degeneracy between the reddened-color and redshift is still unresolved at 3.6 $\mu$m.}

They are not detected by the Kerl G.~Jansky Very Large Array (JVLA) C band (5 cm) deep observation ($\sigma\simeq$ 0.35 $\mu$Jy beam$^{-1}$; \citealt{rujopakarn2016}; Rujopakarn et al.~in preparation). As suggested by \citet{carilli1999}, the flux ratio between radio and (sub-)millimeter wavelengths can be a redshift indicator. In the left panel of Figure \ref{fig:radio-mm_redshift}, we show the redshift dependence of the flux ratio at radio and millimeter wavelengths ($S_\mathrm{5cm}/S_\mathrm{1.2 mm}$). We show the upper limits of the flux ratio of $K$-dropout ASAGAO {sources including 3 candidates}. As comparisons, we also plot the redshift dependence of $S_\mathrm{5cm}/S_\mathrm{1.2 mm}$ of IR bright sources. The result suggests that their flux ratio are roughly consistent with the estimated redshifts (i.e., $z\gtrsim$ 3--5) when we assume the average SED of ALESS sources. In Table \ref{tab:Kdrop_source}, we show the estimated lower limits of photometric redshifts in this case.

\begin{figure*}[ht!]
\plotone{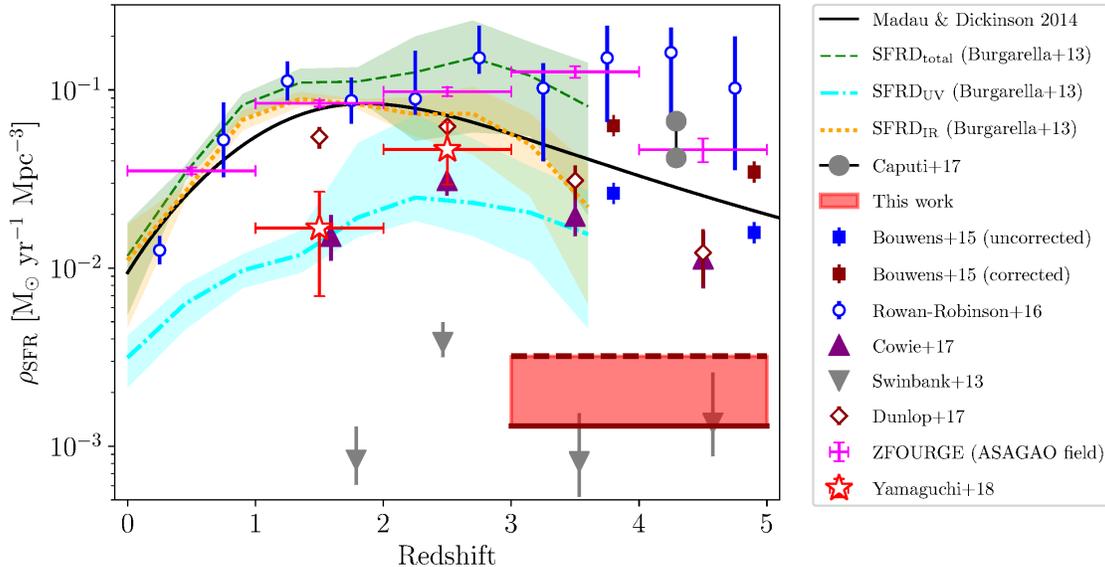}
\caption{{Contribution of ASAGAO sources to the cosmic SFRD as a function of redshift. The red shaded area indicates the contribution of $K$-drop ASAGAO sources. {The horizontal solid line corresponds to the SFR density computed by two secure $K$-dropout ASAGAO sources. The dashed line in the red shaded region indicate the range of SFRD when remaining 3 candidates are also real, respectively (see Section \ref{sec:sfrd} for details).} Red open symbols and magenta symbols are the contributions of ASAGAO sources with $K$-band counterparts and ALMA non-detected ZFOURGE sources within ASAGAO field respectively (Yamaguchi et al.~submitted to ApJ).} The black solid line indicate the recent results of the redshift evolution of the cosmic SFRD obtained by \citet{madau2014}. The green dashed line, cyan dot-dashed line, and orange dotted line show the total (i.e., UV + IR) SFRD, UV SFRD, and IR SFRD obtained by \citet{burgarella2013}. Blue and brown squares are dust-uncorrected and -corrected SFRD obtained by \citet{bouwens2015}. {Blue open circles are results of \citet{rowan2016}. {Purple triangles indicate the cosmic SFRD obtained by the SCUBA2 large survey \citet{cowie2017}. Gray inverse-triangle are the contribution of bright ALESS sources \citep{swinbank2014}.} {Gray filled circles are lower and upper limit of the contribution of H$\alpha$ emitters obtained by \citet{caputi2017}.} Brown open diamonds indicate the contribution of the ALMA sources obtained by \citet{dunlop2017}.} We note that these results are converted to the Chabrier IMF. 
\label{fig:sfrd}}
\end{figure*}

{In the high-redshift case {(i.e., $z\sim$ 4)}, the 3$\sigma$ upper limits of stellar masses of $K$-dropout ASAGAO sources are estimated to be $\log(M_*/M_\odot)$ $\lesssim$ 10.4 using \textit{Spitzer}/IRAC 8.0 $\mu$m, i.e., rest-frame $H$-band data \citep[3$\sigma$ limiting magnitude of 24.3 mag;][]{dickinson2003} if they lie at $z\sim$ 4. Here, we assume a mass-to-light ratio obtained in the rest-frame $H$-band luminosity \citep[e.g.,][]{hainline2011}. \citet{hainline2011} estimated the mass-to-light ratio of dusty sources $M_*/L_H$ = 0.17 and 0.13 $M_\odot\ L_\odot^{-1}$ for constant and single-burst star formation histories, respectively. In this paper, we adopt the average value (i.e., $M_*/L_H$ = 0.15 $M_\odot\ L_\odot^{-1}$) of those two extreme case. We also estimate its IR luminosity by integrating the SED presented in Figure \ref{fig:SED_no_counterpart} and find $\log(L_\mathrm{IR}/L_\odot)\sim$ 12.0 corresponding to $\log(\mathrm{SFR/[M_\odot\ yr^{-1}]})\sim$ 2. {Here, we assume the average SED of ALESS sources at $z\sim$ 4. When we consider the $M_*$--SFR relation at $z\sim$ 4 \citep[e.g.,][]{schreiber2017}, they show starburst-like features.} As discussed in \cite{wang2016}, this source can represent the early phase of formation of massive galaxies, which are difficult to be observed using rest-frame ultraviolet (UV) selected galaxies such as Lyman-$\alpha$ emitters or Lyman break galaxies.} 

{In the low-redshift case {(i.e., $z\sim$ 2)}, {we can estimate the stellar mass of ID17 to be   $\log(M_*/M_\odot)$ $\simeq$ 9.4, because it is detected at \textit{Spitzer}/IRAC 4.5 $\mu$m data, which delivers the rest-frame $H$-band light at $z\sim$2.} According to \citet{straatman2016}, a completeness limit of the ZFOURGE survey is $\log(M_*/M_\odot)$ $\sim$ 9.0 at $z$ = 2, which implies that ID17 prefers the high-redshift case rather than the low-redshift case. For other 4 $K$-dropout ASAGAO {sources including 3 candidates}, their 3$\sigma$ upper limits of stellar masses are estimated to be $\log(M_*/M_\odot)$ $\lesssim$ 8.8 when we consider the 3$\sigma$ limiting magnitude of S-CANDELS (26.5 mag)\footnote{{For ID20, it is difficult to use \textit{Spitzer}/IRAC photometries because of heavy confusions (Figure \ref{fig:stamp_no_counterpart})}}. These upper limits are consistent with their non-detections at $K$-band. Thus, we can not exclude the low-redshift case for these 4 sources. If they lie at $z\sim$ 2, their IR luminosities are estimated to be $\log(L_\mathrm{IR}/L_\odot)\sim$ 11.6 when we assume the SED template of \citet{dale2002} with $T_\mathrm{dust}$ = 25 K. Therefore, in this case, they seem to be extremely low-mass starburst galaxies, which have been missed in previous deep surveys at optical/near-IR wavelengths.}

\section{Contribution to the cosmic SFRD}
\label{sec:sfrd}

Many previous studies predict that the contribution of dust-obscured star-forming activities to the cosmic SFRD have a peak level at $z\simeq$ 2--3 and decline toward $z\gtrsim$ 3--4 based on, for example, IR luminosity functions obtained by the \textit{Herschel} \citep[e.g.,][]{burgarella2013} or dust attenuation-corrected UV observations \citep[e.g.,][]{bouwens2015}. On the other hand, \citet{rowan2016} predict that the contribution seems to be constant at $z$ = 1--5 based on the integrated SFR functions estimated by \textit{Hershel}/SPIRE-500 $\mu$m sources. 

According to \citet{simpson2014}, their optical/near-IR-dark SMGs are located in the redshift range of $z\sim$ 3--5. Thus, in this section, we assume the case that all of $K$-dropout ASAGAO {candidates} lie somewhere in the redshift interval of $z$ $\sim$ 3--5. When we use the average SED of ALESS sources, their contribution to the cosmic IR SFRD is estimated to be $\rho_\mathrm{SFR}\sim$ 1--3 $\times10^{-3}$ $M_\odot\ \mathrm{yr^{-1}\ Mpc^{-3}}${, which corresponding to $\sim$ 10--30\% of previous works \citep[e.g.,][]{madau2014}}. Here, we simply sum up the SFRs of $K$-dropout ASAGAO sources and divide them by the co-moving volume. {The uncertainty of their contributions to the cosmic IR SFRD in Figure \ref{fig:sfrd} are attributable to the relativity of $K$-dropout ASAGAO candidates. If only 2 secure sources with counterparts (i.e., ID17 and ID20) are real, their contribution is expressed by the solid horizontal dark-red line in Figure \ref{fig:sfrd} ($\rho_\mathrm{SFR}\sim$ 1 $\times10^{-3}$ $M_\odot\ \mathrm{yr^{-1}\ Mpc^{-3}}$). On the other hand, in the case that all 5 sources are real, their contribution is shown by the dark-red dashed horizontal line ($\rho_\mathrm{SFR}\sim$ 3 $\times10^{-3}$ $M_\odot\ \mathrm{yr^{-1}\ Mpc^{-3}}$).} 

{We also consider uncertainty attributed to different assumed SEDs. If we estimated SFRs of $K$-dropout ASAGAO candidates using SED templates presented in Figure \ref{fig:radio-mm_redshift}, their contributions to the cosmic IR SFRD can vary by $\sim$ $\pm$ 0.3 dex, which dose not affect following our conclusion significantly.}

As shown in Figure \ref{fig:sfrd}, their contributions to the cosmic SFRD can be comparable with, or greater than that of bright ALESS SMGs ($S_\mathrm{870\mu m}$ $>$ 4.2 mJy; \citealt{swinbank2014}). Therefore, the non-negligible contribution of dust-obscured star formation activities to the cosmic SFRD at high redshift could have been missed in previous surveys. This result shows the importance of ALMA deep contiguous survey to study the evolution of the cosmic SFRD.

\acknowledgements
{We thank the referee for the comments, which improved the manuscript.}
This paper makes use of the following ALMA data: ADS/JAO.ALMA\#2015.1.00098.S, 2015.1.00543.S, and 2012.1.00173.S. ALMA is a partnership of ESO (representing its member states), NSF (USA), and NINS (Japan) together with NRC (Canada), NSC and ASIAA (Taiwan), and KASI (Republic of Korea) in cooperation with the Republic of Chile. The Joint ALMA Observatory is operated by ESO, AUI/NRAO, and NAOJ.
Data analysis was partly carried out on the common-use data analysis computer system at the Astronomy Data Center (ADC) of the National Astronomical Observatory of Japan.
Y.~Yamaguchi is thankful for the JSPS fellowship.
This study was supported by the JSPS Grant-in-Aid for Scientific Research (S) JP17H06130 and the NAOJ ALMA Scientific Research Number 2017-06B.


\begin{thebibliography}{}
\bibitem[Aravena et al.(2016)]{aravena2016} 
Aravena, M., Decarli, R., Walter, F., et al.\ 2016, \apj, 833, 68 
\bibitem[Ashby et al.(2015)]{ashby2015} 
Ashby, M.~L.~N., Willner, S.~P., Fazio, G.~G., et al.\ 2015, \apjs, 218, 33 
\bibitem[Bouwens et al.(2013)]{bouwens2013} 
Bouwens, R.~J., Oesch, P.~A., Illingworth, G.~D., et al.\ 2013, \apjl, 765, L16 
\bibitem[Bouwens et al.(2015)]{bouwens2015} 
Bouwens, R.~J., Illingworth, G.~D., Oesch, P.~A., et al.\ 2015, \apj, 803, 34 
\bibitem[Burgarella et al.(2013)]{burgarella2013} 
Burgarella, D., Buat, V., Gruppioni, C., et al.\ 2013, \aap, 554, A70 
\bibitem[Caputi et al.(2012)]{caputi2012} 
Caputi, K.~I., Dunlop, J.~S., McLure, R.~J., et al.\ 2012, \apjl, 750, L20 
\bibitem[Caputi et al.(2017)]{caputi2017} 
Caputi, K.~I., Deshmukh, S., Ashby, M.~L.~N., et al.\ 2017, \apj, 849, 45 
\bibitem[Carilli \& Yun(1999)]{carilli1999} 
Carilli, C.~L., \& Yun, M.~S.\ 1999, \apjl, 513, L13 
\bibitem[Casey et al.(2018)]{casey2018} 
Casey, C.~M., Hodge, J., Zavala, J.~A., et al.\ 2018, \apj, 862, 78 
\bibitem[Chabrier(2003)]{chabrier2003} 
Chabrier, G.\ 2003, \pasp, 115, 763 
\bibitem[Cowie et al.(2017)]{cowie2017} 
Cowie, L.~L., Barger, A.~J., Hsu, L.-Y., et al.\ 2017, \apj, 837, 139 
\bibitem[Cowie et al.(2018)]{cowie2018}
Cowie, L.~L., Gonz{\'a}lez-L{\'o}pez, J., Barger, A.~J., et al.\ 2018, \apj, 865, 106 
\bibitem[da Cunha et al.(2015)]{dacunha2015} 
da Cunha, E., Walter, F., Smail, I.~R., et al.\ 2015, \apj, 806, 110
\bibitem[Dale \& Helou(2002)]{dale2002} 
Dale, D.~A., \& Helou, G.\ 2002, \apj, 576, 159 
\bibitem[Dickinson et al.(2003)]{dickinson2003} 
Dickinson, M., Giavalisco, M., \& GOODS Team 2003, The Mass of Galaxies at Low and High Redshift, 324 
\bibitem[Downes et al.(1986)]{downes1986} 
Downes, A.~J.~B., Peacock, J.~A., Savage, A., \& Carrie, D.~R.\ 1986, \mnras, 218, 31 
\bibitem[Dunlop et al.(2017)]{dunlop2017} 
Dunlop, J.~S., McLure, R.~J., Biggs, A.~D., et al.\ 2017, \mnras, 466, 861 
\bibitem[Fazio et al.(2004)]{fazio2004} 
Fazio, G.~G., Hora, J.~L., Allen, L.~E., et al.\ 2004, \apjs, 154, 10 
\bibitem[Ford et al.(1998)]{ford1998} 
Ford, H.~C., Bartko, F., Bely, P.~Y., et al.\ 1998, \procspie, 3356, 234 
\bibitem[Franco et al.(2018)]{franco2018} 
Franco, M., Elbaz, D., B{\'e}thermin, M., et al.\ 2018, \aap, 620, A152 
\bibitem[Fujimoto et al.(2016)]{fujimoto2016} 
Fujimoto, S., Ouchi, M., Ono, Y., et al.\ 2016, \apjs, 222, 1 
\bibitem[Fujimoto et al.(2018)]{fujimoto2018} 
Fujimoto, S., Ouchi, M., Kohno, K., et al.\ 2018, \apj, 861, 7 
\bibitem[Gaia Collaboration et al.(2016)]{gaia2016} 
Gaia Collaboration, Brown, A.~G.~A., Vallenari, A., et al.\ 2016, \aap, 595, A2 
\bibitem[Grogin et al.(2011)]{grogin2011} 
Grogin, N.~A., Kocevski, D.~D., Faber, S.~M., et al.\ 2011, \apjs, 197, 35 
\bibitem[Hainline et al.(2011)]{hainline2011} 
Hainline, L.~J., Blain, A.~W., Smail, I., et al.\ 2011, \apj, 740, 96 
\bibitem[Hatsukade et al.(2016)]{hatsukade2016} 
Hatsukade, B., Kohno, K., Umehata, H., et al.\ 2016, \pasj, 68, 36 
\bibitem[Hatsukade et al.(2018)]{hatsukade2018} 
Hatsukade, B., Kohno, K., Yamaguchi, Y., et al.\ 2018, \pasj, 70, 105
\bibitem[Hodge et al.(2013)]{hodge2013} 
Hodge, J.~A., Karim, A., Smail, I., et al.\ 2013, \apj, 768, 91 
\bibitem[Kimble et al.(2008)]{kimble2008} 
Kimble, R.~A., MacKenty, J.~W., O'Connell, R.~W., \& Townsend, J.~A.\ 2008, \procspie, 7010, 70101E 
\bibitem[Koekemoer et al.(2011)]{koekemoer2011} 
Koekemoer, A.~M., Faber, S.~M., Ferguson, H.~C., et al.\ 2011, \apjs, 197, 36 
\bibitem[Kohno et al.(2016)]{kohno2016} 
Kohno, K., Yamaguchi, Y., Tamura, Y., et al.\ 2016, Galaxies at High Redshift and Their Evolution Over Cosmic Time, 319, 92 
\bibitem[Madau \& Dickinson(2014)]{madau2014} 
Madau, P., \& Dickinson, M.\ 2014, \araa, 52, 415 
\bibitem[Rowan-Robinson et al.(2016)]{rowan2016} 
Rowan-Robinson, M., Oliver, S., Wang, L., et al.\ 2016, \mnras, 461, 1100 
\bibitem[Rujopakarn et al.(2016)]{rujopakarn2016} 
Rujopakarn, W., Dunlop, J.~S., Rieke, G.~H., et al.\ 2016, \apj, 833, 12.
\bibitem[Schreiber et al.(2017)]{schreiber2017} 
Schreiber, C., Pannella, M., Leiton, R., et al.\ 2017, \aap, 599, A134.
\bibitem[Silva et al.(1998)]{silva1998} 
Silva, L., Granato, G.~L., Bressan, A., \& Danese, L.\ 1998, \apj, 509, 103 
\bibitem[Simpson et al.(2014)]{simpson2014} 
Simpson, J.~M., Swinbank, A.~M., Smail, I., et al.\ 2014, \apj, 788, 125 
\bibitem[Simpson et al.(2015)]{simpson2015} 
Simpson, J.~M., Smail, I., Swinbank, A.~M., et al.\ 2015, \apj, 807, 128 
\bibitem[Skelton et al.(2014)]{skelton2014} 
Skelton, R.~E., Whitaker, K.~E., Momcheva, I.~G., et al.\ 2014, \apjs, 214, 24 
\bibitem[Straatman et al.(2016)]{straatman2016} 
Straatman, C.~M.~S., Spitler, L.~R., Quadri, R.~F., et al.\ 2016, \apj, 830, 51
\bibitem[Swinbank et al.(2014)]{swinbank2014} 
Swinbank, A.~M., Simpson, J.~M., Smail, I., et al.\ 2014, \mnras, 438, 1267
\bibitem[Tadaki et al.(2015)]{tadaki2015} 
Tadaki, K.-i., Kohno, K., Kodama, T., et al.\ 2015, \apjl, 811, L3 
\bibitem[Tamura et al.(2010)]{tamura2010} 
Tamura, Y., Iono, D., Wilner, D.~J., et al.\ 2010, \apj, 724, 1270 
\bibitem[Ueda et al.(2018)]{ueda2018} 
Ueda, Y., Hatsukade, B., Kohno, K., et al.\ 2018, \apj, 853, 24 
\bibitem[Vega et al.(2008)]{vega2008} 
Vega, O., Clemens, M.~S., Bressan, A., et al.\ 2008, \aap, 484, 631 
\bibitem[Wang et al.(2016)]{wang2016} 
Wang, T., Elbaz, D., Schreiber, C., et al.\ 2016, \apj, 816, 84 
\bibitem[Wang et al.(2009)]{wang2009} 
Wang, W.-H., Barger, A.~J., \& Cowie, L.~L.\ 2009, \apj, 690, 319 
\bibitem[Wang et al.(2016)]{wangwh2016} 
Wang, W.-H., Kohno, K., Hatsukade, B., et al.\ 2016, \apj, 833, 195 
\bibitem[Yamaguchi et al.(2016)]{yamaguchi2016} 
Yamaguchi, Y., Tamura, Y., Kohno, K., et al.\ 2016, \pasj, 68, 82 
\bibitem[Yun et al.(2008)]{yun2008} 
Yun, M.~S., Aretxaga, I., Ashby, M.~L.~N., et al.\ 2008, \mnras, 389, 333 

\end{thebibliography}
\end{document}